\documentclass[conference]{IEEEtran}

\usepackage[pdftex]{graphicx}

\usepackage{scalerel}
\usepackage{tikz}
\usetikzlibrary{svg.path}
\definecolor{orcidlogocol}{HTML}{A6CE39}
\tikzset{
  orcidlogo/.pic={
    \fill[orcidlogocol] svg{M256,128c0,70.7-57.3,128-128,128C57.3,256,0,198.7,0,128C0,57.3,57.3,0,128,0C198.7,0,256,57.3,256,128z};
    \fill[white] svg{M86.3,186.2H70.9V79.1h15.4v48.4V186.2z}
                 svg{M108.9,79.1h41.6c39.6,0,57,28.3,57,53.6c0,27.5-21.5,53.6-56.8,53.6h-41.8V79.1z M124.3,172.4h24.5c34.9,0,42.9-26.5,42.9-39.7c0-21.5-13.7-39.7-43.7-39.7h-23.7V172.4z}
                 svg{M88.7,56.8c0,5.5-4.5,10.1-10.1,10.1c-5.6,0-10.1-4.6-10.1-10.1c0-5.6,4.5-10.1,10.1-10.1C84.2,46.7,88.7,51.3,88.7,56.8z};
  }
}
\newcommand\orcidicon[1]{\href{https://orcid.org/#1}{\mbox{\scalerel*{
\begin{tikzpicture}[yscale=-1,transform shape]
\pic{orcidlogo};
\end{tikzpicture}
}{|}}}}

\usepackage{url}
\usepackage[utf8]{inputenc}
\usepackage{textcomp}
\usepackage{gensymb}
\usepackage[T1]{fontenc}
\usepackage{hyperref}
\hypersetup{
	colorlinks,
	citecolor=black,
	filecolor=black,
	linkcolor=black,
	urlcolor=black
}

\begin{document}

\title{A Generic Bundle Forwarding Interface}

\author{
\IEEEauthorblockN{Felix Walter \orcidicon{0000-0002-4724-8092}}
\IEEEauthorblockA{D3TN GmbH\\        
	Dresden, Germany\\
	<firstname>.<lastname>@d3tn.com}
}

\maketitle

\begin{abstract}
    A generic interface for determining the next hop(s) for a DTN bundle is a valuable contribution to DTN research and development as it decouples the topology-independent elements of bundle processing from the topology-dependent forwarding decision.
    We introduce a concept that greatly increases flexibility regarding the evaluation and deployment of DTN forwarding and routing techniques and facilitates the development of software stacks applicable to heterogeneous topologies.
\end{abstract}

\IEEEpeerreviewmaketitle

\section{Motivation}\label{sec:introduction}

A universal deployment of \emph{Delay- and Disruption-tolerant Networking (DTN)} will consist of a huge number of diverse, heterogeneous (sub-)networks, such as networks with:
\begin{itemize}
	\item \textbf{persistent low-latency end-to-end connectivity},
	\item short or long-latency, asymmetric or unidirectional, but \textbf{precisely-scheduled \emph{contacts}} between the nodes,
	\item \textbf{probabilistic encounters} (e.g., if nodes move somewhat randomly but stay in local vicinity most of the time),
	\item or totally \textbf{random connectivity}.
\end{itemize}
Further variations of these characteristics can be imagined.
To prevent the need to develop specialized implementations of the DTN protocols for every (sub-)network, we advocate for a flexible interface attached to common core infrastructure.

%%%%%%%%%%%

\section{Generic Interface Concept}\label{sec:concept}

\begin{figure}[ht]
	\centering
	\includegraphics[width=.98\linewidth]{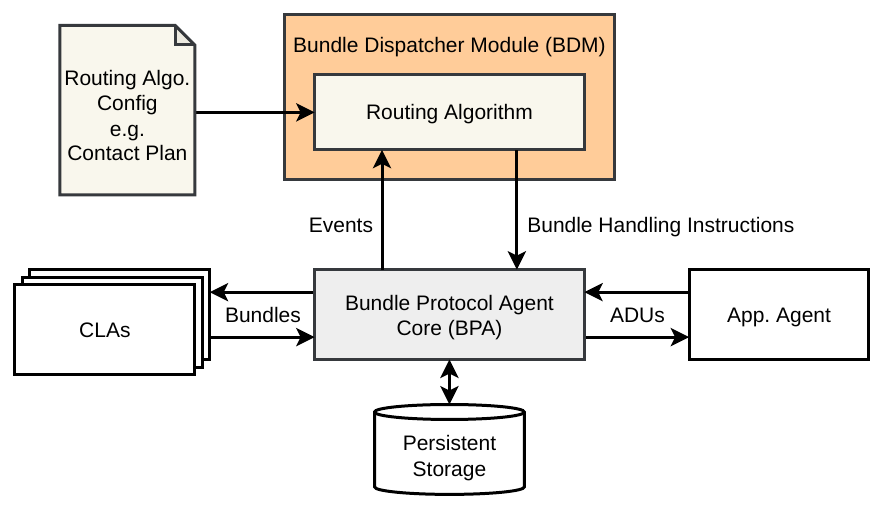}
	\caption{High-level overview of the Bundle Dispatcher Module concept as extension to Figure 2 from the DTN Bundle Protocol specification, RFC 9171 \cite{rfc9171}. The concept will be implemented in µD3TN \cite{impl:ud3tn}.}
	\label{fig:concept-high-level}
\end{figure}

As sketched in Figure \ref{fig:concept-high-level}, our concept introduces a dedicated component integrating with the \emph{Bundle Protocol Agent (BPA)}, which we call the \emph{Bundle Dispatcher Module (BDM)}.
Communication occurs via a network or IPC socket using a low-overhead data exchange format with wide compatibility such as Protobuf or Cap'n Proto.

\subsection{Event-based Bundle Processing}

To achieve a loose coupling of the components and increased flexibility with respect to the BDM implementation, the interface leverages an event-based approach:
The BPA Core posts information about changes in connectivity, bundles that require a forwarding decision, bundles expiring, etc. via a \emph{publish-subscribe} interface.
A module subscribing to these \emph{events} can then decide a) whether or not to act on them and b) invoke behavior to realize the forwarding decision.
This approach further provides the flexibility to attach multiple \emph{subscribers} beside the BDM, e.g., for monitoring purposes.

To pass information about bundles to the BDM, a \emph{Bundle Metadata} data structure is defined, containing the bundle headers, extension blocks, and further metadata like timestamps, but excluding the payload for performance reasons.

\subsection{Bundle Processing Actions}

As counterpart to the event-based information flow to the BDM, a \emph{Remote Procedure Call (RPC)} mechanism is provided by the BPA Core, which offers functions for updating the stored bundles and associated forwarding decisions.
In this context, we propose to attach a list of \emph{Bundle Processing Actions} to each bundle, that defines in-order what should be done next with the given bundle and can be updated by the BDM.
At least two actions must be supported:
\begin{itemize}
    \item \textbf{SendTo(node)}: Forward the bundle to the specified next-hop node. Note that this action may occur multiple times or contain a multicast identifier.
    \item \textbf{Drop}: Remove the ``Forward pending'' retention constraint if the previous action was successful.
\end{itemize}

This list can get implementation-dependent extensions, e.g., actions to fragment bundles, control their storage, and so on.
Like in some Software-Defined Networking implementations, there might be a function to query the supported actions for announcing such extensions in a straight-forward manner.

The BPA Core executes the action lists beginning with the bundle that was updated first, which allows the BDM to control the order of bundle forwarding.
It should be noted that an empty action list is a valid state: In this case, the BPA Core will keep the bundle in memory until its lifetime expires or the BDM triggers an update on the action list.
For maximum flexibility, the action list can have a configurable default assigned to bundles upon reception, thus, even allowing BPA operation without a dedicated BDM in simple cases (e.g., if there is a persistent ``default gateway'' contact).

\subsection{BDM Interaction}

\begin{figure}[ht]
	\centering
	\includegraphics[width=.92\linewidth]{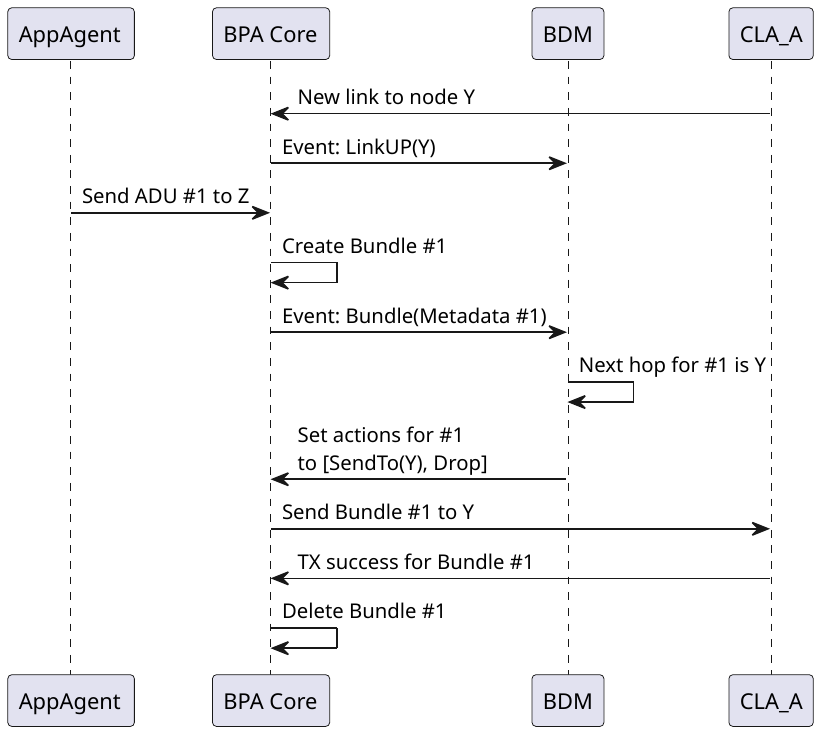}
	\caption{Example sequence diagram for injecting a new Application Data Unit (ADU) addressed to node Z for which the next hop is node Y, assuming opportunistic single-copy forwarding in the BDM (simplified).}
	\label{fig:seq-inject}
\end{figure}

\begin{figure}[ht]
	\centering
	\includegraphics[width=.98\linewidth]{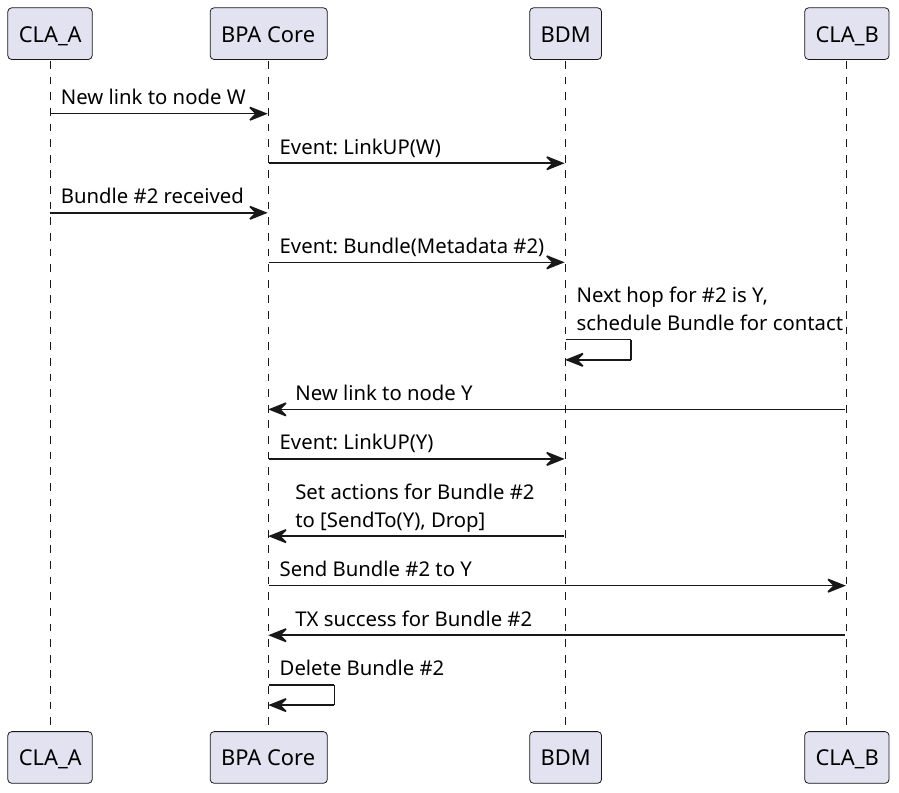}
	\caption{Example sequence diagram for receiving and forwarding a bundle addressed to node Z for which the next hop is node Y, assuming contact-based routing in the BDM (simplified).}
	\label{fig:seq-forward}
\end{figure}

The two complementary mechanisms provide for a flexible interaction between BPA Core and BDM and the implementation of basically arbitrary forwarding techniques.
Figures \ref{fig:seq-inject} and \ref{fig:seq-forward} depict how two example cases would be handled.

\vspace{-5pt} % reduce spacing between figures

\subsection{Inter-BDM Communication}

Some forwarding techniques require the exchange of specific data, e.g., PRoPHET \cite{dtn:prophet} needs to distribute delivery predictability values between nodes.
A BDM implementation has three options to exchange such data: a) attach it to bundles as extension blocks (the used metadata structure contains them), b) inject bundles on its own by registering as a BP application, or c) use an independent mechanism or channel.

%%%%%%%%%%%

\section{Related Work}\label{sec:related}

The presented concept enables a DTN software stack that can adapt to heterogeneous topologies, for which some alternative approaches exist.
\emph{DTN2} \cite{impl:dtn2} implements bundle forwarding in dedicated C++ classes that can be triggered by over 60 different events and can access the bundle data structure in memory.
The \emph{Interplanetary Overlay Network (ION)} \cite{impl:ion} consists of multiple daemons that access a shared data structure. The forwarding decision is made in a separate daemon, which is chosen depending on the destination endpoint scheme.
\emph{ProgDTN} \cite{dtn7-progdtn} is a novel approach executing a JavaScript program from within the BPA that represents a function returning the forwarding decision and can be flexibly exchanged.
Overall, however, none of these implementations provides as much flexibility to implement and exchange the forwarding component as the approach introduced here.

%%%%%%%%%%%

\section{Summary and Outlook}\label{sec:conclusion}

This paper outlined a clean, low-overhead, event-driven socket interface to support a dedicated bundle forwarding component that can be flexibly exchanged and, thus, provides adaptability to heterogeneous topologies.
At the time of writing, a combined Rust (of the BPA Core) and Python 3 (of the BDM) implementation in µD3TN \cite{impl:ud3tn} is almost complete and the integration of techniques such as Schedule-Aware Bundle Routing (SABR) \cite{dtn:sabr} and IP-based Neighbor Discovery (IPND) \cite{dtn:ipnd} is ongoing.
The pending evaluation of this interface will be based on an extended field test using drones and a satellite link to emulate a heterogeneous DTN internetwork.

%%%%%%%%%%%

\section*{Acknowledgments}

The presented concept stems from a long series of discussions in which most of the team at D3TN was involved.
Specifically, without the extensive contributions of the following people, the concept would not be as advanced: Marius Feldmann, Juan Andres Fraire, Tobias Nöthlich, and Georg Alexander Murzik.
The concept is developed as part of the REDMARS2 project that is funded by Germany's Federal Ministry of Education and Research (FKZ16KIS1356).

%%%%%%%%%%%

\end{document}